\documentclass[a4paper]{article}
\pdfoutput=1
\usepackage{natbib}

\usepackage[english]{babel}
\usepackage[utf8x]{inputenc}
\usepackage{amsmath}
\usepackage{color}
\usepackage{url}
\usepackage{graphicx}

\begin{document}
\begin{flushleft}
{\Large
\textbf\newline{A Bayes Factor for Replications of ANOVA Results}
}
\newline
\\
Christopher Harms\textsuperscript{1,*}
\\
\bigskip
\textsuperscript{\bf{1}} Rheinische Friedrich-Wilhelms-Universität Bonn, Germany \\
\textsuperscript{*} Email: christopher.harms@uni-bonn.de
\\
\bigskip
\textbf{This pre-print is equivalent to the author accepted version of the article. Please cite as:} Harms, C. (in press). A Bayes Factor for Replications of ANOVA Results. \emph{The American Statistician}.
\\
\bigskip

\end{flushleft}

\section*{Abstract}
With an increasing number of replication studies performed in psychological science, the question of how to evaluate the outcome of a replication attempt deserves careful consideration. Bayesian approaches allow to incorporate uncertainty and prior information into the analysis of the replication attempt by their design. The Replication Bayes factor, introduced by \cite{Verhagen2014}, provides quantitative, relative evidence in favor or against a successful replication. In previous work by \citet{Verhagen2014} it was limited to the case of $t$-tests. In this paper, the Replication Bayes factor is extended to $F$-tests in multi-group, fixed-effect ANOVA designs. Simulations and examples are presented to facilitate the understanding and to demonstrate the usefulness of this approach. Finally, the Replication Bayes factor is compared to other Bayesian and frequentist approaches and discussed in the context of replication attempts. R code to calculate Replication Bayes factors and to reproduce the examples in the paper is available at \url{https://osf.io/jv39h/}.

\section{Introduction}
The ''replication crisis'' \citep{Pashler2012,Asendorpf2013} has been a focus of discussion in psychological science in recent years. There is ongoing debate about how to improve methodological rigor and statistical analysis in order to improve reliability and replicability of published findings. When considering a given replication study, a central question is the final evaluation, i.e. the question whether a previous finding has successfully been replicated.

As with any scientific question in empirical disciplines, researchers use statistical tools in order to find an answer to this question. An intuitive way to evaluate replication results is to compare statistical significance in both an original study and a replication study (''vote counting''): If both studies are significant and show effects in the same direction, the replication is deemed ''successful''. If, on the other hand, the original study reported a statistically significant test which is non-significant in the replication study, the replication might be considered ''failed''. This interpretation, while intuitive and common practice in past research \citep{Maxwell2015}, is flawed in general and wrong in the case of non-significant replications. A non-significant result cannot be interpreted as evidence for the absence of an effect -- especially since the difference between a significant result in an original study and a non-significant result in a replication study might not be significant in itself \citep{Gelman2006}. Moreover, statistical significance and $p$-values do not contain information about uncertainty in estimates and their misinterpretations have repeatedly been covered over the past decades \citep{Wasserstein2016,Ionides2017,Nickerson2000,Bakan1966}. A related question is whether the strict dichotomy between ''successful'' and ''failed'' replications is sensible in practice.

To overcome problems with simply comparing $p$-values based on their significance, other methods for evaluating replication studies have been proposed, partly also alleviating the strict dichotomy: confidence intervals (CI) for effect size estimates have been advocated by some authors for both the reporting of statistical summaries \citep{Cumming2001,Cumming2012} and for the evaluation of replications (\citealp{Cumming2008}; \citealp{Gilbert2016}, but see also \citealp{Anderson2016b}). A researcher could, for example, check if the effect size estimate from a replication study falls within the 95\% confidence interval of the effect size estimate from the original study -- or, \emph{vice versa}, check if the effect size estimate from the original study falls within the 95\% confidence interval of the replication study. While this approach takes uncertainty into account (in contrast to $p$-values), confidence intervals can be difficult to interpret and give rise to misinterpretations \citep{Belia2005,Anderson2016b}.

The width of a confidence interval is directly related to sample size and thus to the power curve of a statistical test. Since power is notoriously low in psychological science \citep{Button2013,Sedlmeier1989,Szucs2017}, confidence intervals tend to be generally wide for original studies from the published literature \citep{morey2016most}. This can make a sensible comparison of confidence intervals difficult and lead to inconclusive or misleading results (depending on the particular decision rule). \citet{Simonsohn2015} therefore proposed a method to take the power of the original study into account when evaluating a replication study. In his ''small telescopes'' approach, the confidence interval of a replication study is compared not to the effect size estimate from the original study, but to an effect the original study had 33\% power to detect. The benchmark of 33\% is arbitrary, but \cite{Simonsohn2015} considers it a ''small effect''. The goal is to determine if a replication study, yielding a smaller effect size than in the original study, is still in line with a small effect that the original study had only little power to detect. This allows for a more nuanced interpretation of a replication outcome than simply comparing $p$-values or confidence intervals.

The approaches outlined so far rely on a frequentist interpretation of probability and data analysis. Frequentist statistics is primarily concerned with repeated sampling and long-run rates of events. It does not allow a researcher to answer questions such as ''How much more should I believe in an effect?'' or ''Given the data, what are the most credible effect size estimates?''. Bayesian statistics allows to address such questions in a coherent framework and uses uncertainty in estimates as an integral part of the analysis.

Related to the analysis of a replication study, various Bayesian approaches are possible. \cite{Marsman2017}, for example, have evaluated the results from the ''Reproducibility Project: Psychology'' \citep{OpenScienceCollaboration2015} using several methods, including Bayesian parameter estimation on the individual study level, Bayesian meta-analysis, and Bayes factors. For psychology in particular, Bayes factors \citep{Kass1995,Jeffreys1961} have repeatedly been proposed as an alternative or addition to significance testing in common study scenarios (see \citealp{Morey2016}, or \citealp{Bayarri2015}, for introductions to Bayesian hypothesis testing).

\cite{Verhagen2014} compared several Bayes factors to be used in the context of replication studies and introduced a Bayes factor to specifically investigate the outcome of a replication study, which they termed \emph{Replication Bayes factor}. It is constructed to be used only with reported test-statistics from an original study and a replication study. As some researchers disagree with the notion of ''belief'' in data-analysis, it is also a favorable property of the Replication Bayes factor that it makes only little assumptions about prior beliefs. The Replication Bayes factor allows to test the hypothesis that the effect in a replication study is in line with an original study against the hypothesis that there is no effect. The result is a continuous quantification of relative evidence \citep{Morey2016}, showing by much the data are more in line with one hypothesis compared to the other.

The present article focuses on the Bayesian perspective and extends the Replication Bayes factor. \citet{Verhagen2014} introduced the Replication Bayes factor solely for the case of one- and two-sample $t$-tests. In this paper, it will be extended to the case of $F$-tests in fixed-effect ANOVA designs, a common study design in cognitive, social, and other sub-fields of psychology. In order to do so, an outline of hypothesis testing using Bayes factors and how it can be applied to replication studies is presented. In the second section, it is shown how the Replication Bayes factor can be used in studies investigating difference between several groups, when fixed-effect ANOVAs are carried out. Simulations and example studies are then followed by a general discussion of the method.

In general, it has been recommended \citep{Brandt2014,Anderson2016} that replicators should include different methods to evaluate the outcome of a replication study to account for the advantages, limitations, and varying statistical questions the different approaches present. This allows for a more nuanced interpretation of replication studies than a simple comparison of statistical significance. The Replication Bayes factor is hence presented as an addition to existing methods of analyzing the outcome of a replication study. The article aims to provide an accessible introduction in the setup of the Replication Bayes factors, but assumes a general understanding of Bayesian statistics -- in particular with the general concepts of priors, marginal likelihoods, and Bayesian belief updating. Readers unfamiliar with the Bayesian approach might find the excellent textbooks by \citet{McElreath2016} and \citet{Kruschke2015} helpful. A reading list of relevant introductory papers is given by \citet{Etz2015b}. \cite{Rouder2012a} and \cite{Morey2016} have elaborated specifically on the use of Bayes factors.

To reproduce the examples in this article, all scripts are available in an OSF repository at \url{https://osf.io/jv39h/} \citep{Harms_2018_OSF}. To calculate the Replication Bayes factor for $t$-tests \citep{Verhagen2014} and $F$-tests (the present article), an R-Package is available at \url{https://github.com/neurotroph/ReplicationBF}.

\section{Bayesian Hypothesis Testing}
While Bayes' theorem allows several interpretations, Bayes factors are best understood in the context of \emph{Bayesian belief updating}. Bayes factors indicate how to rationally shift beliefs in two competing hypotheses (formalized as models $M_0$ and $M_1$) based on the observed data $Y$:
\begin{equation}
	\label{eq:bayesian-updating}
	\underbrace{\frac{P(M_0 | Y)}{P(M_1 | Y)}}_{\text{Posterior Odds}} =
	 \underbrace{\frac{\pi({M_0})}{\pi(M_1)}}_{\text{Prior Odds}} \times
	 \underbrace{\frac{P(Y | M_0)}{P(Y | M_1)}}_{\text{Bayes factor}}
\end{equation}

The Bayes factor $BF_{01} = \frac{P(Y | M_0)}{P(Y | M_1)}$ is the factor by which the prior odds are multiplied in order to get updated posterior odds \citep{Jeffreys1961,Lindley1993,Kass1995}. It is a ratio of the \emph{marginal likelihoods} or \emph{model evidence} of the two models, which is given by
\begin{equation}
	\label{eq:marginal-likelihood}
	P(Y | M_i) = \int p(Y | \theta, M_i) \pi(\theta | M_i) \ \mathrm{d}\theta
\end{equation}
where $\theta$ is the vector of model parameters, $\pi(\theta | M_i)$ is the prior distribution for the parameters in Model $M_i$ and $p(Y | \theta, M_i)$ is the likelihood function of $M_i$. The marginal likelihood is the normalizing constant in Bayes' rule in order for the posterior to be a proper probability distribution. Hence, the integrand consists of the same parts as the nominator in Bayes' rule, namely likelihood and prior.

The challenge of computing Bayes factors lies in the possible complexity of the integral. As will be explained in more detail later, in many practical cases -- such as in the case for the Replication Bayes factor -- the marginal likelihood needs to be approximated, for example by using Monte Carlo methods.

Bayes factors provide relative evidence for one model when compared to another model and they can -- in contrast to $p$-values in the Neyman-Pearson framework of null-hypothesis significance testing -- be interpreted directly in a continuous, quantitative manner. To facilitate verbal interpretation, \cite{Jeffreys1961} and \cite{Kass1995} provided guidelines for the description of Bayes factors: A $BF_{10}$ is ''not worth more than a bare mention'' if it is between 1 (the data provide evidence for both models equally) and about 3; the evidence against model $M_0$ is ''substantial'' if $3 < BF_{10} \leq 10$, ''strong'' if $10 < BF_{10} \leq 100$ and ''decisive'' if $BF_{10} > 100$.

In recent years it has become increasingly popular to report Bayes factors as an addition or alternative to traditional null-hypothesis significance testing using $p$-values. Calculations and tools for common scenarios in psychological research exist \citep[e.g.][]{Rouder2012a,Rouder2012b,Rouder2009,Dienes2016}. In the context of replications, \cite{Etz2016} have used Bayes factors to show that the original studies in the ''Reproducibility Project: Psychology'' \citep{OpenScienceCollaboration2015} have provided only little relative evidence against the null hypothesis (when taking into account publication bias). Replication studies yielding substantial in favor or against the null hypothesis were mostly studies with larger sample sizes. The analysis is an example how Bayes factors could be used to provide more insights into the results of empirical studies.

To summarize, the Bayesian frameworks allows a researcher to incorporate existing, previous knowledge into the statistical analysis. This is particularly useful in the context of replication studies: An original experiment has provided information about an effect and if one wants to evaluate a replication study, the information from the original should be considered. The Replication Bayes factor formalizes this and tests the hypothesis of a successful replication against the hypothesis of a null effect.

\subsection{Bayes factors for Replications}
The \emph{Replication Bayes factor} introduced by \cite{Verhagen2014} is a way to use the Bayesian hypothesis testing framework in the context of replications and to quantify the outcome of a replication attempt given the information from the original study. The general idea is to use the posterior from the original study as a prior for the analysis of the replication attempt -- in line with the idea of updating beliefs as outlined above. Furthermore, it is desirable to only use data that is easily available from published studies. Thus, ideally, only reported test-statistics with degrees of freedom and sample size are used.

For a Bayesian hypothesis test, two models need to be set up and compared in the Bayes factor. For the Replication Bayes factor these two models or hypotheses about an effect size measure $\delta$ are:
\begin{enumerate}
	\item $H_0: \delta = 0$, that is the hypothesis that the true effect size is zero (the ''skeptic's position'' in the terms of \citealp{Verhagen2014}).
	\item $H_r: \delta \approx \delta_{\mathrm{orig}}$, i.e. the hypothesis that the original study is a good estimate of the true effect size (the ''proponent's position'').
\end{enumerate}

The Replication Bayes factor is then the ratio of the marginal likelihoods of the two models considering the data from the replication study, denoted by $Y_{\mathrm{rep}}$ \citep[p. 1461]{Verhagen2014}:
\begin{equation}
	\label{eq:bf-rep-general}
	\text{B}_{\text{r}0} = \frac{p( Y_{\mathrm{rep}} | H_r )}{p( Y_{\mathrm{rep}} | H_0 )}
\end{equation}

That is, the Replication Bayes factor represents the relative evidence of the data in favor of a successful replication when compared to a null hypothesis of no effect. ''Successful'' here means, that the replication yields a similar effect size as the original study. This implies an assumption about the validity of the effect size estimate of the original study, which is also discussed later in this paper.

In order to calculate $\text{B}_{\text{r}0}$ one needs to mathematically define the two models and find a useful representation of the data from the original study ($Y_{\mathrm{orig}}$) and the replication study ($Y_{\mathrm{rep}}$). While the data would ideally be the raw data and used to fit a probabilistic model describing the relationship between independent and dependent variables, for the Replication Bayes factor it is desired to rely on more accessible data. In the case of (published) replication studies often only summary and test statistics are reported. Thus, for the evaluation of replication studies it is desirable to be able to calculate the Replication Bayes factor based on the reported test statistics (with degrees of freedom and sample sizes) from the original study and the replication study alone.

\subsection{Replication Bayes factor for $t$-tests}
In their paper, \citet{Verhagen2014} provided models and formulas for the computation of the Replication Bayes factor for $t$-tests. In the following section, the rationale and the setup of the model will be reiterated as a basis for the extension to $F$-tests.

The general model is based on the $t$-distribution used for the $t$-test and derives from the distribution of $t$-values under the alternative hypothesis. The marginal likelihood for the two models that are compared is given
\begin{equation}
	\label{eq:model-evidence-t}
	p(Y_{\mathrm{rep}} | H_i) = \int t_{df_{\mathrm{rep}}, \delta \sqrt{N_{\mathrm{rep}}}}(t_{\mathrm{rep}}) \pi(\delta | H_i)\ \mathrm{d}\delta
\end{equation}
with $t_{df, \Delta}(x)$ being the non-central $t$-distribution with $df$ degrees of freedom and non-centrality parameter $\Delta = \delta \sqrt{N_{\mathrm{rep}}}$.

The two models under consideration differ in their prior distributions $\pi(\delta | H_i)$: For the skeptic ($H_0$), the prior distribution is 1 at $\delta = 0$ and 0 everywhere else. Thus, the marginal likelihood simplifies to
\begin{equation}
	\label{eq:marginal-likelihood-skeptic-t}
	p(Y_{\mathrm{rep}} | H_0) = t_{df_{\mathrm{rep}}}(t_{\mathrm{rep}})
\end{equation}
Which is the central $t$-distribution (i.e. non-central $t$-distribution with non-centrality parameter $\Delta = 0$) evaluated at the point of the $t$-value observed in the replication study, $t_{\mathrm{rep}}$.

For the proponent on the other hand, the prior distribution for the replication, $\pi(\delta | H_r)$, is the posterior distribution of the original study. If one starts out with a flat, uninformative prior before the original experiment, the resulting posterior distribution was described as $\Lambda^\prime$-distribution by \cite{Lecoutre1999}. While in Bayesian statistics flat priors are often disregarded in favor of at least minimally regularizing priors \citep{Gelman2013}, in the present case the prior for the original study plays a minor rule: It is quickly overruled by the data even when considering only a single original study. For the Replication Bayes factor, only the posterior of the original study is relevant. The $\Lambda^\prime$-distribution can be approximated closely through a normal distribution as \cite{Verhagen2014} showed in their appendix. If the prior for the replication, $\pi(\delta | H_r)$, is expressed through the posterior of the original study, so that $\pi(\delta | H_r) = p(\delta | \delta_{\mathrm{orig}}, H_r)$, the marginal likelihood for the proponent's model is finally given as
\begin{equation}
	\label{eq:marginal-likelihood-proponent-t}
	p(Y_{\mathrm{rep}} | H_r) = \int t_{df_{\mathrm{rep}}, \delta \sqrt{N_{\mathrm{rep}}}}(t_{\mathrm{rep}}) p(\delta | \delta_{\mathrm{orig}}, H_r) \ \mathrm{d}\delta
\end{equation}

For the integral no closed form is yet known. Hence, it needs to be approximated and different methods exist. \cite{Verhagen2014} used the Monte Carlo estimate \citep[][chap. 7.2.1]{Gamerman2006}: random samples from the posterior-turned-prior distribution, $p(\delta | \delta_{\mathrm{orig}}, H_r)$, are repeatedly drawn and the average of the marginal likelihood term is calculated. Other approaches to approximating the marginal likelihood for the Bayes factor are available and will be discussed below.

Putting Equations~\ref{eq:marginal-likelihood-skeptic-t} and \ref{eq:marginal-likelihood-proponent-t} into Equation~\ref{eq:bf-rep-general} yields the formula for the Replication Bayes factor for $t$-tests:
\begin{equation}
	\label{eq:bf-rep-t}
	\text{B}_{\text{r}0} = \frac{\int t_{df_{\mathrm{rep}}, \delta \sqrt{N_{\mathrm{rep}}}}(t_{\mathrm{rep}}) p(\delta | \delta_{\mathrm{orig}}, H_r) \ \mathrm{d}\delta}{t_{df_{\mathrm{rep}}}(t_{\mathrm{rep}})}
\end{equation}

In their paper, \cite{Verhagen2014} used simulation studies and examples to demonstrate the usefulness of the Replication Bayes factor, and show how it compares to other Bayes factors that might also be used for replication studies.

\section{Replication Bayes factor for $F$-tests}
Many studies in psychological research do not compare two independent groups through $t$-tests, but investigate differences across multiple groups and interactions between factors. Thus, there seems to be a need to extend this approach to other tests such as the $F$-test in ANOVAs. In this section the steps necessary to apply the Replication Bayes factor to other tests are explained, and the Bayes factor is derived for the $F$-test in fixed-effect ANOVA designs.

As a general difference to the $t$-test, $F$-tests do not convey information about the direction or location of an effect; the hypothesis under investigation is an omnibus hypothesis if the effect degrees of freedom are $df_{\mathrm{effect}} > 1$ \citep{rosenthal2000contrasts,Steiger2004}. Therefore incorporating the $F$-statistic in the Replication Bayes factor does not allow researchers to take the direction of the effect into account. As will be shown in Example 3 and discussed later, researchers need to consider additional information when evaluating the outcome of replication studies with ANOVA designs. Nevertheless, researchers use and report $F$-tests both for omnibus and interaction hypotheses. And the $F$-statistic does contain information about the effect size. Thus the $F$-statistic can be used in the Replication Bayes factor if a statement about the size of an effect is desired.

In order to maintain the general nature of the Replication Bayes factor from Equation~\ref{eq:bf-rep-general}, an effect size measure needs to be chosen to parametrise the model. Cohen's $f^2$ \citep{Cohen1988} has a simple relationship to the non-centrality parameter $\lambda$ of the non-central $F$-distribution \citep{Steiger2004}:

\begin{equation*}
	\lambda = f^2 \cdot N
\end{equation*}
Since this relationship holds only in the case of fixed-effects models, the Replication Bayes factor for $F$-tests can also only be used validly in these cases -- a limitation that will also be discussed later.

Setting up the model based on the $F$-distribution in a way similar to the $t$-test case (cf. Equation~\ref{eq:model-evidence-t}) leads to the marginal likelihoods
\begin{equation}
	\label{eq:model-evidence-F}
	p(Y_{\mathrm{rep}} | H_i) = \int F_{df_{\mathrm{effect}},df_{\mathrm{error}},\lambda}(F_{\mathrm{rep}}) \pi(f^2 | H_i) \ \mathrm{d}f^2
\end{equation}
Where $F_{df_{\mathrm{effect}},df_{\mathrm{error}},\lambda}(x)$ is the noncentral $F$-distribution with degrees of freedom $df_{\mathrm{effect}}$ and $df_{\mathrm{error}}$ and noncentrality parameter $\lambda$.

For the skeptic's position, $H_0$, the marginal likelihood simplifies -- analogous to the $t$-test case -- to the central $F$-distribution evaluated at the observed $F$-value from the replication study, since the prior $\pi(f^2 | H_0)$ is chosen so that it is 1 at $f^2 = 0$ and 0 everywhere else:
\begin{equation}
	\label{eq:marginal-likelihood-skeptic-F}
	p(Y_{\mathrm{rep}} | H_0) = F_{df_{\mathrm{effect}},df_{\mathrm{error}}}(F_{\mathrm{rep}})
\end{equation}

\begin{figure}[tb!]
\centering
\includegraphics[width=1.0\textwidth]{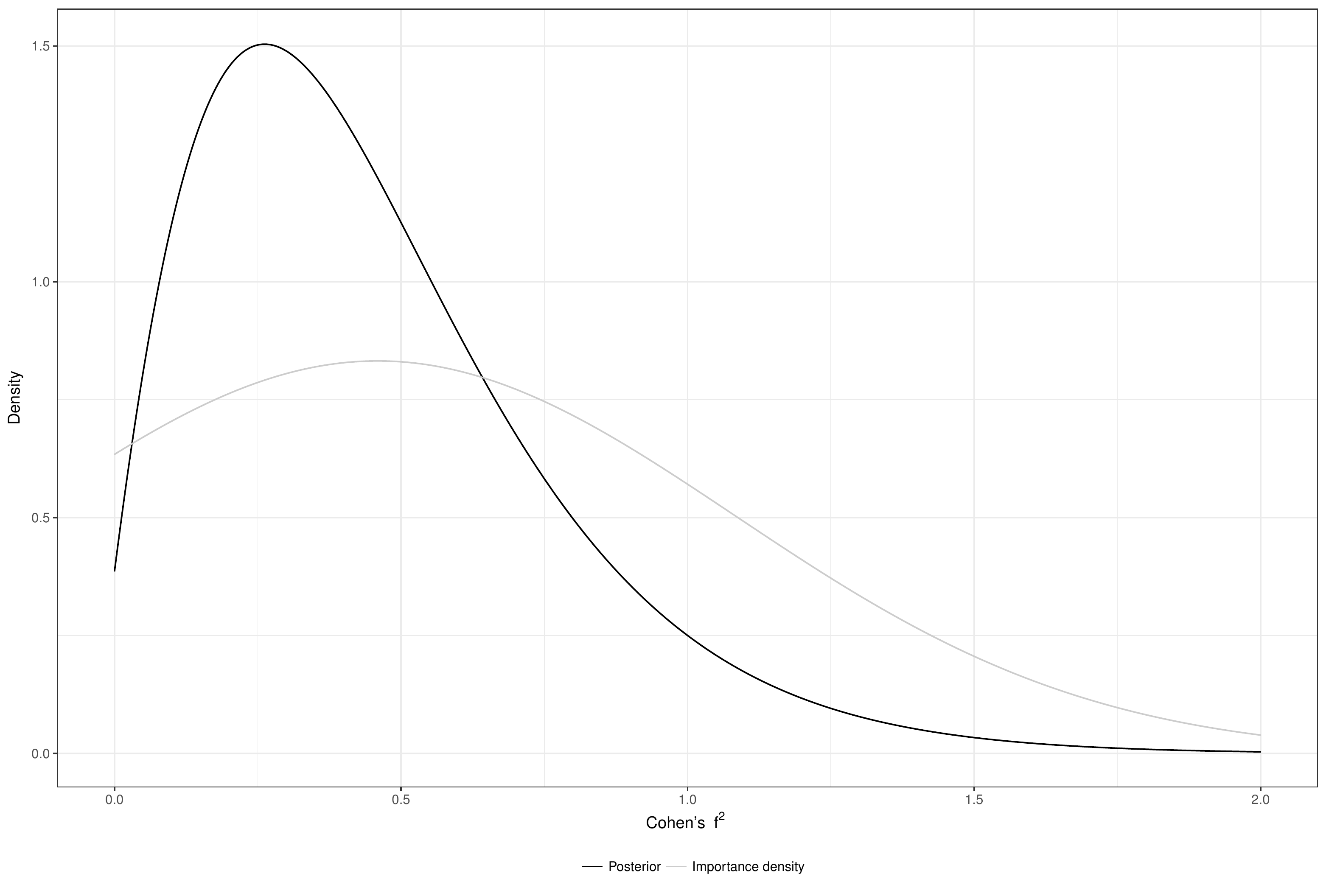}
\caption{\label{fig:normal-approx}Plot of posterior distribution for a study with an observed $f^2 = 0.625$ in an one-way ANOVA with two groups and 10 participants each, i.e. 20 participants in total, yielding $F(1, 8) = 5.0$. The grey line represents the half-normal importance density $g(f^2)$, which is constructed based on samples from the posterior distribution generated by Metropolis-Hastings, used for estimating the marginal likelihood.} 
\end{figure}

For the $H_r$ model on the other hand, the prior $\pi(f^2 | H_r)$ should again be the posterior of the original study. Starting out with a uniform prior before the original study results in the following posterior distribution for the original study:
\begin{equation}
	\label{eq:posterior-original-F}
	p(f^2 | Y_{\mathrm{orig}}) = \frac{F_{df_{\mathrm{effect, orig}},df_{\mathrm{error, orig}},f^2\cdot N_{\mathrm{orig}}}(F_{\mathrm{orig}})}{\int F_{df_{\mathrm{effect, orig}},df_{\mathrm{error, orig}},f^2\cdot N_{\mathrm{orig}}}(F_{\mathrm{orig}})\ \mathrm{d}f^2}
\end{equation}
\cite{Lecoutre1999} called this distribution $\Lambda^2$-distribution, similar to the $\Lambda^\prime$-distribution for $t$-tests. In contrast, this distribution cannot be easily approximated by a normal distribution (see shape of distribution in Figure~\ref{fig:normal-approx}). Despite the use of an improper prior, the posterior distribution is valid as all parameters are actually observed in the original study, i.e., $df_{\mathrm{effect, orig}}, df_{\mathrm{error, orig}}, F_{\mathrm{orig}} > 0$ and $N_{\mathrm{orig}} \gg 1$.

The Replication Bayes factor for $F$-tests is then given by
\begin{equation}
	\label{eq:bf-rep-F}
	\text{B}_{\text{r}0} = \frac{\int \! F_{df_{\mathrm{effect}},df_{\mathrm{error}},f^2\cdot N}(F_{\mathrm{rep}}) p(f^2 | Y_{\mathrm{orig}}) \ \mathrm{d}f^2}{F_{df_{\mathrm{effect}},df_{\mathrm{error}}}(F_{\mathrm{rep}})}
\end{equation}

The challenge to compute the Replication Bayes factor lies primarily in the calculation of the integral of the numerator. Marginal likelihoods in general are often difficult to compute or intractable and analytical solutions are rarely available in applied settings. \citet{Verhagen2014} chose the Monte Carlo estimator \citep[p. 239]{Gamerman2006}. For the Monte Carlo estimate samples are randomly drawn from the prior distribution of the $H_r$ model, i.e. from the original study's posterior distribution $p(f^2 | Y_{\mathrm{orig}})$, the integrand is calculated and the marginal likelihood is approximated by taking the average.
\begin{equation}
	\mathrm{B}_{\mathrm{r}0} \approx \frac{1}{M} \sum_{i}^{M} \frac{F_{df_\mathrm{effect}, df_\mathrm{error}, f^2_{(i)}\cdot N_{\mathrm{rep}}}(F_\mathrm{rep})}{F_{df_\mathrm{effect}, df_\mathrm{error}}(F_\mathrm{rep})}, \qquad f^2_{(i)} \sim p(f^2 | Y_\mathrm{orig})
\end{equation}

The Monte Carlo estimate, however, is inefficient and unstable, especially when prior and likelihood disagree (i.e. when the original and the replication study yield very different effect size estimates). There are other estimators for the marginal likelihood available. \citet[chap. 7]{Gamerman2006} provide an overview on twelve different estimators. \citet{Bos2002} compared seven estimators in simulations and showed, that the Monte Carlo estimate is highly unstable and often yields values very different from an analytically derived solution. \emph{Numerical integration} can work well in low-dimensional settings, but does not scale to problems involving multiple parameters or wide ranges of possible parameter values.

For complex models, \emph{bridge sampling} provides a stable and efficient way to estimate the marginal likelihood \citep{Meng1996,Gronau2017}. Since the present case involves only a single parameter and relatively simple likelihood function and posterior distribution, \emph{importance sampling} is a faster, but also efficient and stable way to estimate the marginal likelihood.

The importance sampling estimate for the marginal likelihood \citep[chap. 7]{Gamerman2006} is calculated by drawing $M$ random samples from an importance density $g(f^2)$ and averaging an adjusted likelihood term:
\begin{align}
	p(Y_{\mathrm{rep}} | H_r) &\approx \frac{1}{M} \sum_{i}^{M} \frac{p(Y_{\mathrm{rep}} | \tilde{f^2_i}) \pi(\tilde{f^2_i})}{g(\tilde{f^2_i})} \nonumber \\
	\label{eq:importance-sampling-estimate}
	&\approx \frac{1}{M} \sum_{i}^{M} \frac{p(Y_{\mathrm{rep}} | \tilde{f^2_i}) p(\tilde{f^2_i} | Y_{\mathrm{orig}})}{g(\tilde{f^2_i})}, \\
	\tilde{f^2_i} &\sim g(f^2) \nonumber
\end{align}

Since the prior for $H_r$ is the posterior distribution given in Equation~\ref{eq:posterior-original-F}, the marginal likelihood $p(Y_{\mathrm{orig}} | H_r)$ for the original study also needs to be computed in order for $p(f^2 | Y_{\mathrm{orig}})$ to be a proper probability density and the estimator to yield correct results.

How should the importance density $g(f^2)$ be chosen? A simple and straight forward way is to use a half-normal distribution with mean and standard deviation determined by samples from the posterior. The half-normal distribution is easy for drawing random samples and can easily be calculated at any point. While it is not as close to the true posterior (see Figure~\ref{fig:normal-approx}) as in the $t$-test case, it is sufficient for the purpose of importance sampling since the tails of the half-normal distribution are fatter than the non-normalized posterior density (see requirements for importance densities listed in \citealp{Gronau2017}).

To sample from the unknown and yet non-normalized posterior-distribution Markov Chain Monte Carlo (MCMC) techniques can be used, e.g. the Metropolis-Hastings algorithm \citep{Chib1995}. In Bayesian statistics approximating the posterior distribution is one of the core challenges and thus several implementations for different algorithms are available. Metropolis-Hastings yields samples from a target distribution by taking a random walk through the parameter space. An accessible introduction is given in \citet[chap. 8]{McElreath2016} and a more mathematical description can be found in \citet[chap. 11.2]{Gelman2013}. Software packages such as JAGS or Stan \citep{Gelman2015} are commonly used to draw samples from the posterior distribution of Bayesian models.

The mean and standard deviation of the posterior samples are then used to construct a half-normal distribution as importance density $g(f^2)$. Subsequently, random samples from the importance density are used to estimate the marginal likelihood according to Equation~\ref{eq:importance-sampling-estimate}, especially in regions where the posterior distribution has high probability mass, resulting in a better estimate compared to the Monte Carlo estimate.

For the case of the Replication Bayes factor it will be shown below in simulations, that the differences between the estimates are small except in the most extreme cases. Since it is nonetheless desirable to have accurate and robust estimates, in the remainder of the article the Importance Sampling estimate is used.

Dividing the resulting estimate for the marginal likelihood of $H_r$ by the $H_0$ model evidence (Equation~\ref{eq:bf-rep-F}) allows the calculation of the Replication Bayes factor:
\begin{equation}
	\text{B}_{\text{r}0} = \frac{\displaystyle\int F_{df_\text{effect}, df_\text{error}, f^2\times N}(F_\text{rep}) p(f^2 | Y_\text{orig}) \,\mathrm{d}f^2}{F_{df_\text{effect}, df_\text{error}}(F_\text{rep})}
\end{equation}

The resulting Bayes factor can then be interpreted based on its quantitative value: $\text{B}_{\text{r}0} > 1$ is evidence in favor of the proponent's hypothesis, i.e. evidence in favor of a true effect of a similar size, while a $\text{B}_{\text{r}0} < 1$ is evidence against a true effect of similar size. The more the Bayes factor deviates from 1, the stronger the evidence. It might be helpful to use the commonly used boundaries of 3 and $\frac{1}{3}$ for sufficient evidence: $\frac{1}{3} < \text{B}_{\text{r}0} < 3$ is weak evidence for either hypothesis \citep[p. 432]{Jeffreys1961} and should lead the researcher to collect further data to strengthen the evidence \citep{Schonbrodt2015,Edwards1963}.

The provided R-package contains functions to calculate Replication Bayes factors for both $t$- and $F$-tests using the formulas provided here, relying on importance sampling to estimate marginal likelihoods.

\section{Simulation Studies}
In order to show properties of the Replication Bayes factors, three simulation studies are presented. First, the numerical results in different scenarios are shown to allow comparisons across different study designs and effect sizes. Second, the Monte Carlo estimate and the Importance Sampling estimate for the marginal likelihood are compared. While the differences are small except for the most extreme cases, it is argued that the more robust method is to be preferred. Last, the relationship between $t$- and $F$-values in studies of two independent groups is used to compare the Replication Bayes factor for $t$-tests with the proposed adaptation to the $F$-test.

\subsection{Simulation 1: Behavior in Different Scenarios}
\begin{figure}[tb!]
\centering
\includegraphics[width=1.0\columnwidth]{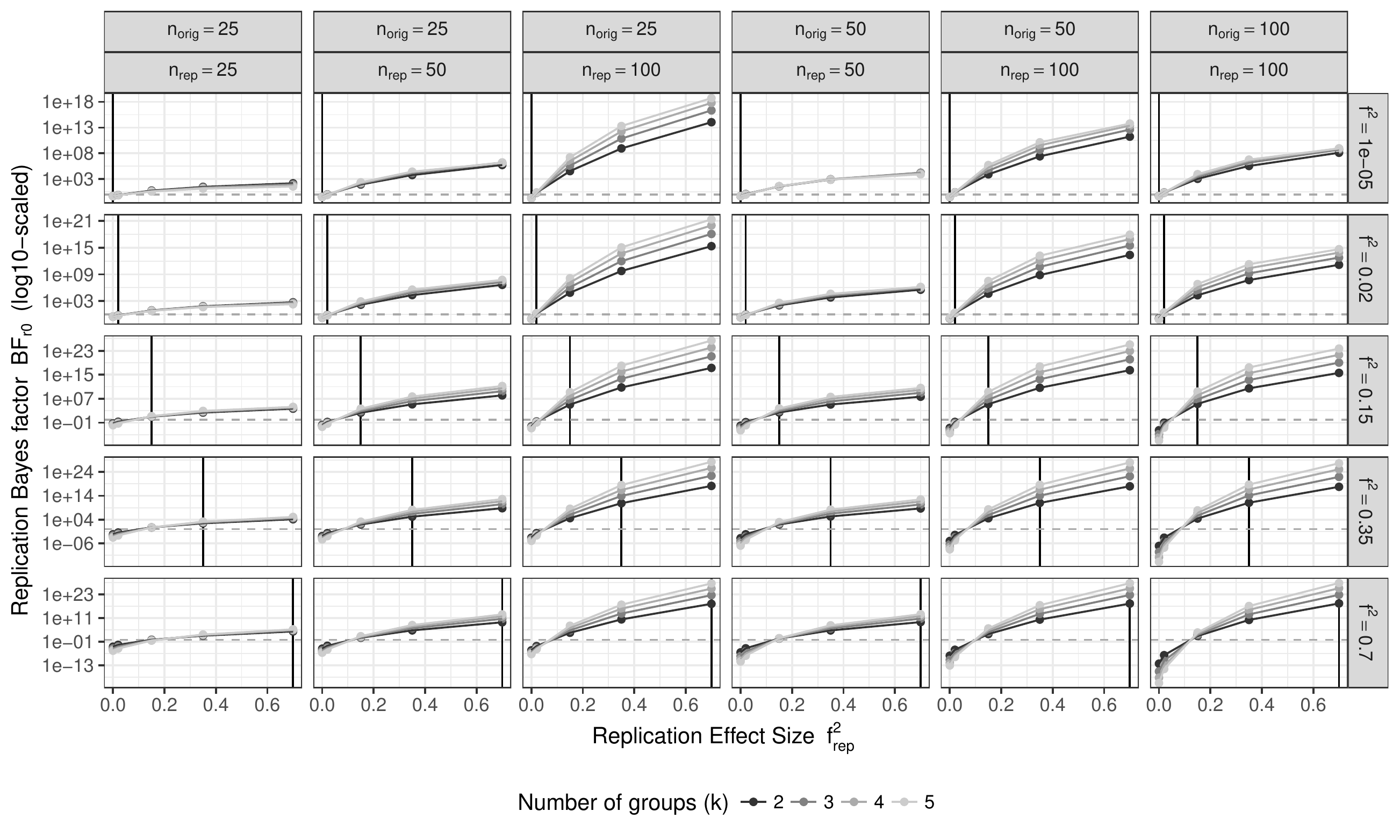}
\caption{\label{fig:bfrep-scenarios}Value of the Replication Bayes factor for $F$-tests in various scenarios. Columns show sample sizes per group in original and replication study, rows are $f^2$ effect sizes in the original study. Horizontal axes in each plot show $f^2$ effect size in replication study and vertical axes are $\log_{10}$-scaled showing $B_{\text{r}0}$. Shades of grey denote number of groups. Each point is one calculated Replication Bayes factor in a given scenario. Vertical black lines indicate the effect size in the original study, i.e. points on this line are replication studies where the effect size equals the original study. Horizontal dashed line is $B_{\text{r}0} = 1$.}
\end{figure}

To better understand the Replication Bayes factor, it is useful to explore different scenarios in which a researcher might calculate it. For Figure~\ref{fig:bfrep-scenarios}, different combinations for original and replication studies were considered. In particular, one can see that the Replication Bayes factor increases towards support for the proponent's position when the replication has large sample size and yields a large effect size estimate.

For situations in which the effect size estimate is very small in both the original and the replication study, the Replication Bayes factor is close to 1. In these cases, the proponent's and the skeptic's position are very similar and even 100 participants per group are not enough to properly distinguish between the two models (see left most points in first row plots of Figure~\ref{fig:bfrep-scenarios}). If, in contrast, the original study reported a relatively large effect and the replication study yields a small effect, the Replication Bayes factor quantifies this correctly as evidence in favor of the skeptic's positions. Moreover, the considered setup shows that -- holding group sizes equal -- more groups allow stronger conclusions since total sample size is higher.

\subsection{Simulation 2: Monte Carlo vs Importance Sampling Estimate}
The aforementioned difference between the Monte Carlo estimate for the marginal likelihood and an importance sampling estimate is especially relevant if prior and likelihood of a model disagree substantially. For the Replication Bayes factor this is the case when original study and replication study yield very different effect size estimates.

As referenced above, \citet{Bos2002} has shown in simulations that the Monte Carlo estimate is generally unstable and leads to biased estimates of the marginal likelihood. He compared the results both against the analytical solution and other estimators, which performed substantially better.

In order to evaluate how this affects the Replication Bayes factor, pairs of studies are investigated. For illustrative purposes, original studies with $n_{\mathrm{orig}} = 15$ per group with an observed effect size of $d_{\mathrm{orig}} \in \{1; 2; 5\}$ and subsequent replications with sample sizes $n_{\mathrm{rep}} \in \{50; 100\}$ per group and observed effect sizes $d_{\mathrm{rep}} \in \{0; 0.3; 0.5\}$ are entered in the Replication Bayes factor for $t$-tests. The resulting Bayes factor is estimated using the Monte Carlo estimate and the Importance Sampling estimate.

\begin{figure}[tbhp]
	\centering
	\includegraphics[width=1.0\columnwidth]{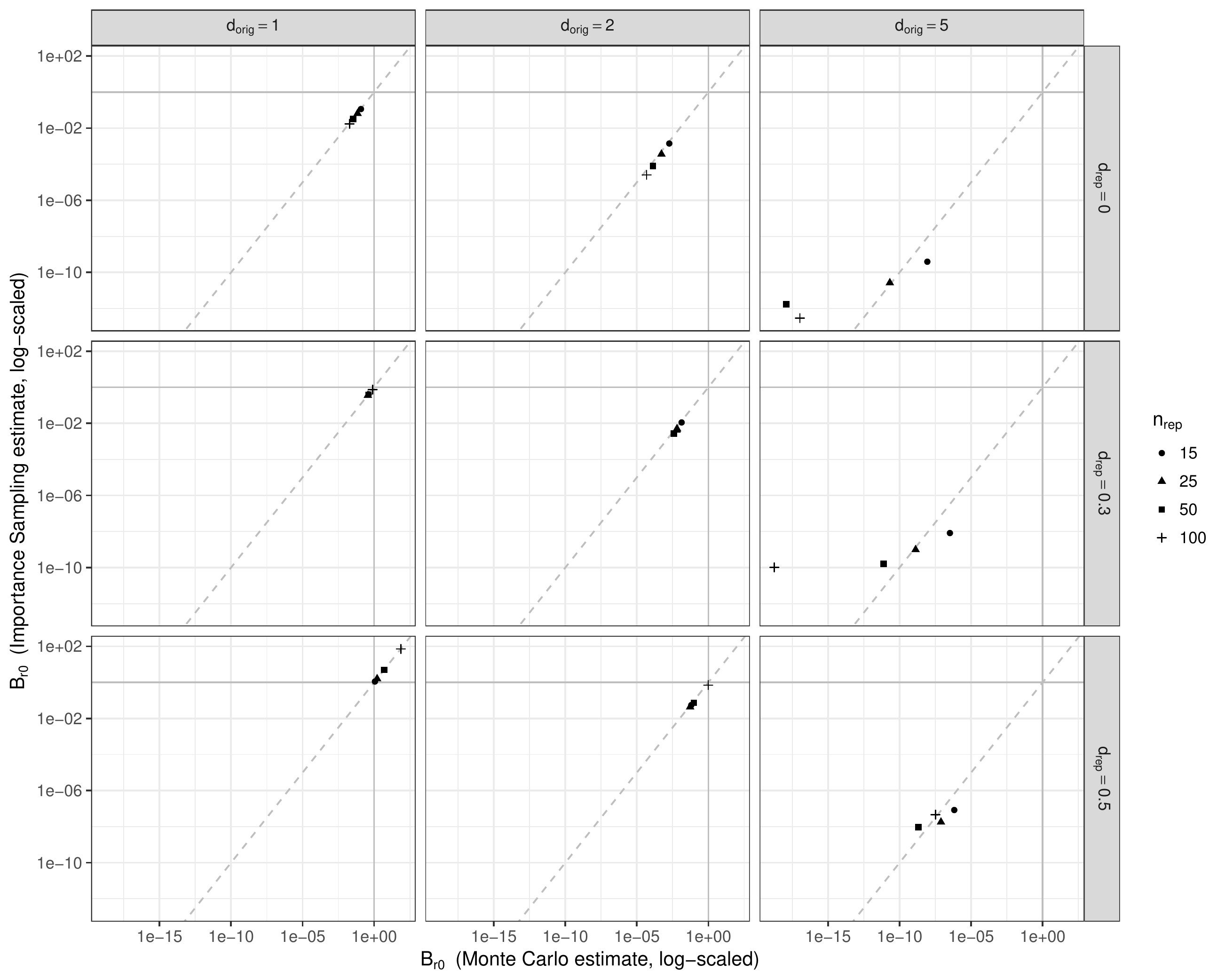}
	\caption{\label{fig:bfrep-sim-estimators}Comparison for Replication Bayes factors ($t$-test) with marginal likelihoods estimated by either Monte Carlo estimation (x-axis) or Importance Sampling estimation (y-axis). Points are individual Bayes factors. Grey dashed line indicates equality between both estimation methods. Only for extreme cases (i.e. $d_{\mathrm{orig}} = 0.5$ and $d_{\mathrm{rep}} > 2$) the estimations yield substantially different results, but conclusions would remain the same. Original studies always had $n_{\mathrm{orig}} = 15$.} 
\end{figure}

As can be seen from the results in Figure~\ref{fig:bfrep-sim-estimators}, the Monte Carlo estimate is generally close to identical to the importance sampling estimate. It is only in extreme cases that there are differences. For example, when an original study yields $d = 5$ in a small sample of $n = 15$ per group and a replication shows only a tiny or even a null effect in a much larger sample. The effect size of $d = 5$ is, however, an implausibly large effect size for most empirical research fields. Even when considering the larger sampling error in small samples, effect sizes around $d = 1$ are already considered a ''large effect'' in social sciences \citep{Cohen1988} -- effects of $d > 3$ do not realistically appear in the literature in psychology.

While the differences can become large in orders of magnitude, in these cases the Replication Bayes factor is also large in general because of the disagreement between original and replication study. Thus, the conclusions are not changed by the difference between estimates. Furthermore, in the practical examples reported below the differences between the two estimates are numerically very small and would not yield different conclusions.

Nevertheless, it should be desirable to provide an accurate estimates of a statistical indicator to make an informed judgement, e.g. about the relative evidence a replication study can provide. The Importance Sampling estimate has been shown to be more stable and more accurate and is thus preferable to the Monte Carlo estimate \citep{Bos2002,Gronau2017,Gamerman2006}.

\subsection{Simulation 3: $F$-test for two groups}
In the context of significance testing, it is a well known relationship that the $F$-tests from a one-way ANOVA with two groups yield the same $p$-values as a two-sided, two-sample $t$-test, when $F = t^2$ is used. Accordingly, it might be an interesting question, how the Replication Bayes factor for $F$-tests relates to the Replication Bayes factor for $t$-tests when used on the same data from two independent groups.

For pairs of original and replication studies with two groups of different sizes ($n_{\mathrm{orig}} \in \{15; 50\}$, $n_{\mathrm{rep}} \in \{15; 30; 50; 100\}$) and different effect sizes ($d_{\mathrm{orig}} \in \{0.2; 0.4; 0.6; 0.8; 1; 2\}$, $d_{\mathrm{rep}} \in \{^1/_{10^{5}}; 0.2; 0.4; 0.6; 0.8; 1; 2\}$), the Replication Bayes factor for $t$-test was calculated using the observed $t$-value and for the $F$-test using $F = t^2$.

\begin{figure}[tbhp]
	\centering
	\includegraphics[width=1.0\columnwidth]{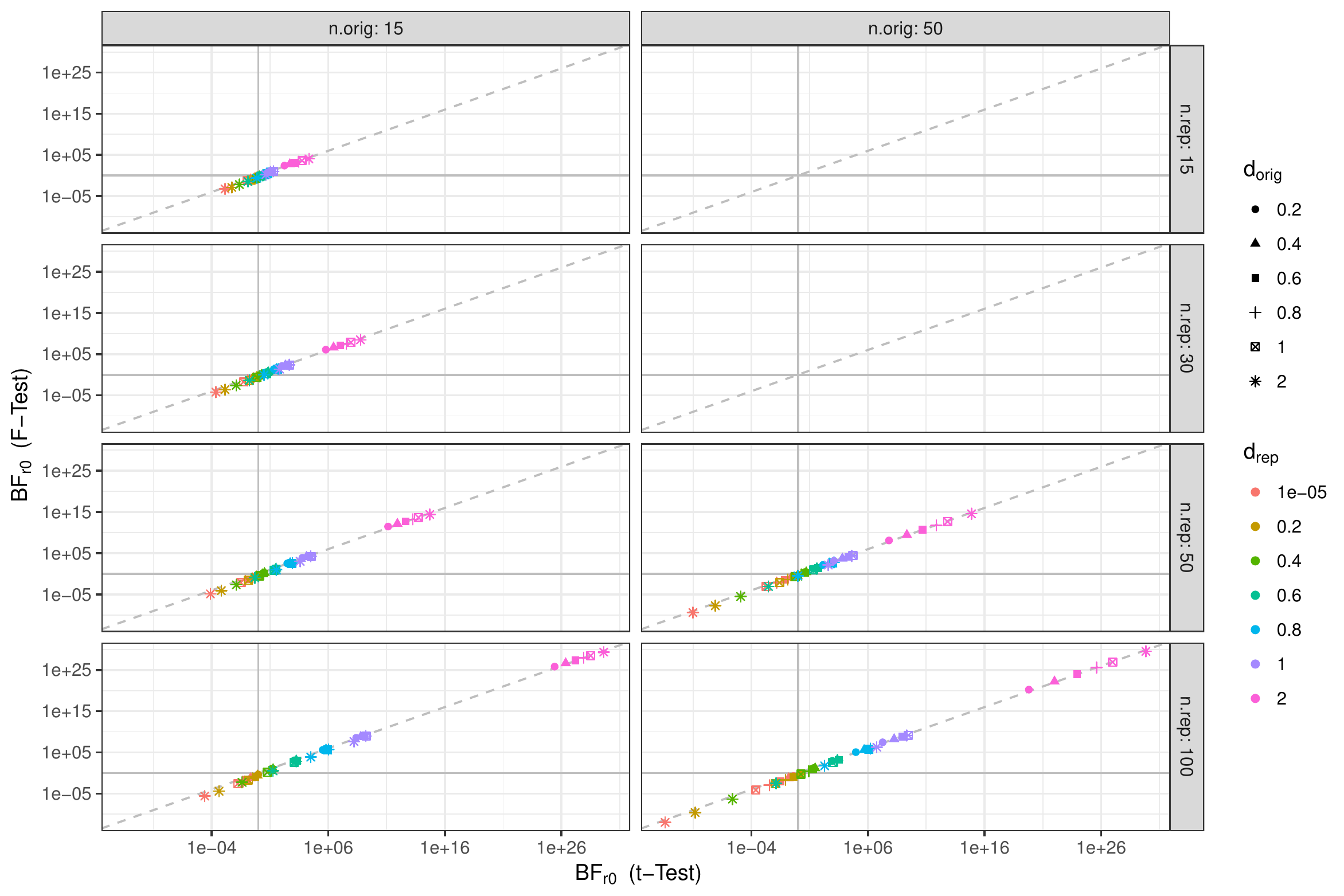}
	\caption{\label{fig:bfrep-sim-tf}Comparison of Replication Bayes factors for $t$- and $F$-test on the same data-set with two groups. Study pairs with replication studies smaller than original studies were not simulated. Dashed grey line is equality.} 
\end{figure}

The results are shown in Figure~\ref{fig:bfrep-sim-tf}. As can be expected, the resulting Bayes factors are very close ($r = 0.999$ across all scenarios). What cannot easily be seen in the figure, however, is that the Replication Bayes factor for the $F$-test is about half the size of the Replication Bayes factor for the $t$-test ($\frac{B_{r0,t}}{B_{r0,F}} = 2.211$).

This result makes intuitively sense: The $F$-statistic does not contain information about the direction of the effect and thus cannot provide the same amount of relative evidence as a Bayes factor on the $t$-test.

\section{Examples}
In this section, the Replication Bayes factor for $F$-tests is applied in two example cases of replication studies from the ''Reproducibility Project: Psychology'' \citep{OpenScienceCollaboration2015}. The third example presented in this section aims to show how to investigate the pattern of effects to ensure valid conclusions based on a high value of the Replication Bayes factor.

\subsection{Example 1}
The first example is an original study conducted by \citet{Albarracin2008}, which was replicated as part of the ''Reproducibility Project: Psychology'' \citep{OpenScienceCollaboration2015} by \citet{Voracek_Sonnleitner_2016} and is available at \url{https://osf.io/tarp4/}.

\citet{Albarracin2008} investigated the effect ''action'' and ''inaction goals'' on subsequent ''motor or cognitive output''. In study 7 specifically, participants were primed with either words relating to ''action'', ''inaction'' or neutral words (control condition). Participants subsequently engaged in either an active or inactive task before they were instructed to read a text and write down their thoughts about the text. The number of listed thoughts was used as a measure for ''cognitive activity''. The experiment thus had a 3 (\emph{Prime}: action, inaction, control) $\times$ 2 (\emph{Task:} active, inactive) between-subjects design.

Participants were predicted to write down more thoughts about the text when they are primed with an ''action'' goal compared to participants primed with an ''inaction'' goal. Futhermore, the possibility to exert activity in an active task, should moderate the effect: ''satisfied action goals should yield less activity than would unsatisfied action goals''. They found the two-way interaction \emph{Prime} $\times$ \emph{Task} to be significant ($F(2, 92) = 4.36$, $p = .02$, $\eta_p^2 = 0.087$ corresponding to $f^2 = 0.095$) in a sample of 98 student participants (group sizes were not reported).

The replication by Sonnleitner and Voracek did not find the same interaction to be significant in their sample of 105 participants, $F(2, 99) = 2.532$, $p = .085$, $\eta_p^2 = 0.049$, $f^2 = 0.051$.

Comparing the replication attempt to the original study using the Replication Bayes factor assuming equally sized groups yields $B_{\mathrm{r}0} = 1.153$, which can be considered inconclusive as the data are nearly equally likely under both models. Based on the $p$-values, it was concluded that the study was not successfully replicated. Judging from the Bayes factor, however, it should be concluded that the replication study did not have enough participants to indicate stronger evidence in either direction. Considering the power of a frequentist test, recent recommendations suggest at least 2.5 times as many participants than in the original study one seeks to replicate \citep{Simonsohn2015}. The Replication Bayes factor thus shows absence of evidence rather than evidence of absence of a failed or successful replication.

For the main effect of \emph{Task}, in contrast, which also was significant in the original study ($F(1, 92) = 7.57$, $p = .007$, $\eta_p^2 = 0.076$, $f^2 = 0.082$) but not in the replication ($F(1, 99) = .107$, $p = .745$, $\eta_p^2 \approx f^2 < .001$), the Replication Bayes factor yields strong evidence in favor of the skeptic's model ($B_{\mathrm{r}0} = 0.057$). This is evidence in favor of an unsuccessful replication of the main effect, since the Bayes factor shows that the data are 17 times more likely under the model implying an effect size $f^2 = 0$ than under the model informed by the original study. It should be noted however, that this does not answer the question whether the replication is in line with an existing albeit smaller effect (which the replication also did not have sufficient power to detect, see \citealp{Simonsohn2015}).

\subsection{Example 2}
For the second example another replication from the ''Reproducibility Project'' is considered: The original study was conducted by \cite{Williams2008} and investigated cues of ''spatial distance on affect and evaluation''. The replication was performed by \citet{Joy-Gaba_Clay_Cleary_2016}. The replication data, materials and final report are available at \url{https://osf.io/vnsqg/}.

In study 4 of the original paper, \cite{Williams2008} have primed 84 participants in three different conditions. The number of participants per group was not reported.  The authors hypothesized that different primes for spatial distance will effect evaluations of perceived ''closeness'' to siblings, parents and hometown. The dependent variable was an index of ratings to those three evaluations. Hence, the study was a between-subjects design with one factor (\emph{Prime}: Closeness, Intermediate, Distance). They found a significant main effect of priming on the ''index of emotional attachment to one's nuclear family and hometown'' in a one-way ANOVA ($F(2, 81) = 4.97$, $p = .009$, $\eta_p^2 = .11$, $f^2 = .124$).

The replication by Joy-Gaba, Clay and Cleary did not find the same main effect in a sample of 125 participants ($F(2, 122) = .24$, $p = .79$, $\eta_p^2 = .003919$, $f^2 = .00393$). Based on the $p$-values they concluded, that the replication was not successful.

But how much more are the data in line with a null model when compared to the proponent's alternative? This is the answer the Replication Bayes factor can give, assuming equal group sizes: $\text{B}_{\text{r}0} = 0.031$. This means, the data is about 32 times more likely under the model stating that the true effect size is 0 than under the model using the original study's posterior.

\subsection{Example 3}
The final example aims to show a caveat when using the Replication Bayes factor for $F$-tests yielding strong support for the proponent's model. As has been addressed above and could be seen in the last simulation, the $F$-statistic (and, consequently, the $f^2$ effect size measure) does not convey information about the location or direction of an effect. This is a general problem when evaluating the outcomes from ANOVA-design studies based on the test-statistic alone. Researchers need additional judgments based on post-hoc $t$-tests or qualitative consideration of interaction plots.

In order to show how to inspect the results thoroughly, an imaginary study is conducted: In an original study, 15 participants each are randomly assigned to three different conditions (45 participants in total). The true population means of the three groups are $\mu_1 = 1.5$, $\mu_2 = 2.2$ and $\mu_3 = 2.9$ and standard deviation is 1 for all groups. Running a one-way ANOVA on a generated data-set yields a significant result, $F(2, 42) = 7.91$, $p = .001$, $f^2 = 0.377$. For the replication study, 30 participants are randomly sampled to the same three conditions each (thus 90 participants in total). For the generated replication data-set the ANOVA yields a significant result as well, $F(2, 87) = 7.60$, $p < .001$, $f^2 = 0.175$.

\begin{figure}[tb!]
	\centering
	\includegraphics[width=1.0\columnwidth]{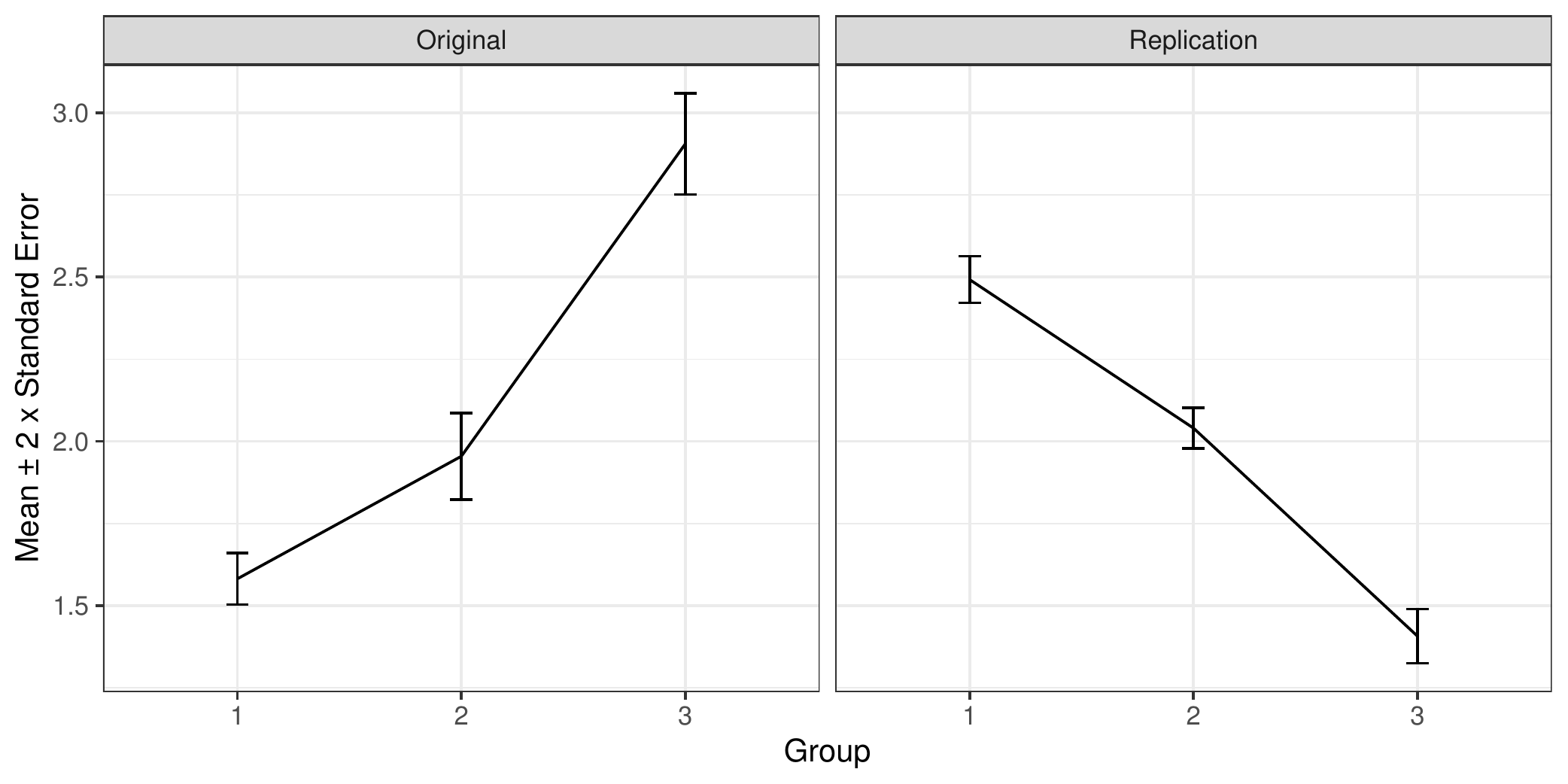}
	\caption{\label{fig:bfrep-ex3-interaction}Summary plot for the imaginary study in Example 3. Original study has 15 participants per group with population means $\mu = (1.5; 2.2; 2.9)$. Replication has 30 participants per group with population means in reverse order.} 
\end{figure}

The test-statistics and effect size estimates show a significant effect in both studies. Did the replication happen to successfully replicate the original finding? Before calculating the Replication Bayes factor, one should inspect the data in more detail. Considering the the plot in Figure~\ref{fig:bfrep-ex3-interaction}, it is obvious that the pattern of effects is strikingly different. In fact, the replication shows the reverse pattern. Since the $F$-statistic is insensitive to this difference, the Replication Bayes factor does not provide us this information. Yet, it does correctly indicate that the data are more in favor of an effect in size -- not direction -- of the original study ($\text{B}_{\text{r}0} = 38.261$).

Considering post-hoc $t$-tests or planned contrasts helps to include location and direction of the effect in the analysis. For the present example, for example, the difference between groups 1 and 3 is significant in both the original study ($t(20.809) = -3.953$, $p = .001$) and the replication study ($t(56.765) = 3.6412$, $p < .001$). The difference in direction is visible in the sign of the $t$-statistic and thus the Replication Bayes factor for $t$-test can consider this information ($\text{B}_{\text{r}0} = 0.0015$ or the reciprocal $\text{B}_{0\text{r}} = 668.15$)

Consequently, researchers investigating $F$-tests should always pay attention to the particular nature of the effect under investigation. This is a recommendation not limited to the evaluation of replication studies, but is generally a crucial step when analyzing data from ANOVA-studies.

\section{Discussion}
The Replication Bayes factor for $F$-tests in fixed-effect ANOVA designs outlined in this paper is an extension of the work by \cite{Verhagen2014}. It utilizes a Bayesian perspective on replications, namely using the results and uncertainty from the original study in the analysis of a replication attempt. The approach outlined in this paper adapts the Replication Bayes factor from $t$- to $F$-tests.

In evaluating replication studies it is reasonable to use the information available from the original study. When $p$-values of two studies are directly compared based on their significance, or when a confidence interval is used to examine whether the effect size in the original can be rejected, this is already essentially done. The Bayesian framework allows to do this in a more formal framework and incorporates uncertainty in estimates. The latter is ignored when only $p$-values are compared. This is one of the reasons why ''vote counting'' is not recommended when evaluating a replication study.

A general criticism of Bayesian hypothesis testing with Bayes factors is the role of the prior. While proponents of Bayesian statistics generally consider incorporating previous knowledge or expectation as a strength of Bayesian statistics, the perceived subjectivity in its selection is troublesome for some non-Bayesians. In contrast to other ways of including the results from the original study in Bayesian model (see below), the Replication Bayes factor introduces very little subjectivity to the analyses. The model for the Replication Bayes factor is derived directly from the assumptions of null-hypothesis significance testing.

While Bayes factors do not provide controls for error rates over many repeated samplings, they do allow for a quantification of relative model evidence. That is, a Bayes factor allows statements such as ''the data are 5-times more likely under the alternative model than under the null model''. Researchers reluctant to the Bayesian approach in general might nevertheless find the Replication Bayes factor in particular a useful addition without adding too much subjectivity. After all, the evaluation of replication studies should not rely on a single method \citep{Marsman2017,Brandt2014,Anderson2016,Gilbert2016}.

The steps taken and explained in this paper can also be used to apply the Replication Bayes factor logic to other tests such as $\chi^2$-tests, $z$-tests, or correlations \citep[see also][Appendix]{Boekel2015}. One needs to use the appropriate test-statistic and sampling distribution under the alternative hypothesis and accordingly derive the resulting marginal likelihoods. The use of Importance Sampling for estimating the marginal likelihoods allows for more general cases than the Monte Carlo estimate; on the one hand it is more robust to large disagreements between original study and replication study (see Simulation 3) and, on the other hand, scales better if more parameters were introduced in the model.

\subsection{Limitations}
As explained above, Cohen's $f^2$ was chosen as the parameter of interest. This limits the application of the proposed Replication Bayes factor for $F$-tests to fixed-effects ANOVAs with approximately equal cell sizes. The relationship between the non-centrality parameter $\lambda$ and $f^2$ is only valid in these cases.

In general, effect size measures in ANOVAs take into account the specific effects (i.e. each cell mean's deviation from the grand mean \citealp{Steiger2004}) and cell sizes. In many reported ANOVAs, however, only the total sample size along with the omnibus or interaction test is reported. Even if specific contrasts are of interest, they are not reported with sample sizes or descriptive statistics required to calculate the specific effects.

What happens when $f^2$ is used in unbalanced designs? If a specific effect is present in a group with larger sample size, $f^2$ will overestimate the overall effect. For studies with randomized allocation to the groups, the differences in cell sizes should be small, so the assumption of balancedness seems warranted. In non-random studies where unbalanced groups are to be expected, the Replication Bayes factor as outlined here should not be used. More elaborate alternatives as suggested below might be more appropriate.

The Replication Bayes factor is also limited by the information provided in the $F$-statistic: Since the $F$-statistic (and consequently $f^2$) does not convey information about the direction and location of the effect, the Replication Bayes factor cannot take the pattern of effects into account. The Replication Bayes factor still does give the relative evidence regarding the size of the effect. As shown in Example 3, one way is to follow-up evidence in favor of a replicated effect by investigating the pattern in the post post-hoc $t$-tests or contrasts. If no such tests are reported, one could use the Replication Bayes factor for the omnibus test as a quantitative indicator and use visual inspection whether the effect is indeed in the same direction. This is in parallel to the recommended practice for analyzing data with ANOVAs.
Another way of dealing with this could be to extend the model used in the Replication Bayes factor to include the descriptive statistics of all groups instead of the test-statistic. This, however, would require additional assumptions and does not follow as directly from the significance tests as the method proposed here. The Bayesian alternatives mentioned below can be a starting point for further extensions of the model.

What is true for $p$-values \citep{Wasserstein2016} also holds true for Bayes factors: No single statistical parameter will take the responsibility from researchers to make a careful evaluation of the different statistical results at hand. Even if all statistical parameters are in agreement, differences in methodology and design might render the statistical comparison \emph{ad absurdum}.

Last, effect size estimates are more difficult to compute for random- and mixed-effects ANOVAs. While effect size measures such as $\omega^2$ can be calculated in these scenarios, the relationship to the non-central $F$-distribution is not as obvious. Further developments of the Replication Bayes factor might be useful in this direction.

\subsection{Alternatives to the Replication Bayes factor}
The Replication Bayes factor is only one way to evaluate a replication study in the Bayesian framework. As outlined above, its strength lies in making little assumptions and the ability to use only reported test-statistic for the evaluation.

One alternative way to use Bayes factors would be to model the information from the original study differently. Where the posterior of the original study is used here as a prior in the replication model, one could also use the effect size estimate and a reported confidence interval to construct a Normal distribution as prior. Statistical software such as JASP \citep{JASP2018} allows to use different priors, including normal distributions, to calculate Bayes factors from test-statistics.

Another way to use the Bayesian framework would be not to rely on Bayes factors but use Bayesian estimation in hierarchical models and to incorporate both the original and the replication study to estimate the effect size across both studies. \citet{Etz2016} have used such models to evaluate the outcomes of the ''Reproducibility Project: Psychology''. Some of the methods outlined in this article can also be used for single study pairs.

Recently, \citet{Ly2017} have proposed a reconceptualization of the Replication Bayes factor. The goal is to avoid the computations for the posterior-turned-prior distribution and instead rely on ''evidence updating''. Different Bayes factors can be multiplied yielding the Replication Bayes factor $\mathrm{BF}_{10}(d_{\mathrm{rep}} | d_{\mathrm{orig}})$ \citep[p. 7]{Ly2017}:

\begin{equation*}
	\mathrm{BF}_{10}(d_{\mathrm{rep}} | d_{\mathrm{orig}}) = \frac{\mathrm{BF}_{10}(d_{\mathrm{orig}}, d_{\mathrm{rep}})}{\mathrm{BF}_{10}(d_{\mathrm{orig}})}
\end{equation*}

This requires the computation of a Bayes factor for the original study alone, $\mathrm{BF}_{10}(d_{\mathrm{orig}})$, and a Bayes factor of a combined data set from both studies, $\mathrm{BF}_{10}(d_{\mathrm{orig}}, d_{\mathrm{rep}})$. In their pre-print, \citet{Ly2017} outline how to calculate this Bayes factor from ''Evidence Updating'' for $t$-tests and contingency tables.

The calculation of an $F$-value representing a combined data-set requires reported means, standard deviations and group sizes for both the original and replication study. These are not always reported and the Replication Bayes factor outlined here is designed to not require it. Therefore the extension of the original concept for the Replication Bayes factor \citep{Verhagen2014} is useful when only limited data is available from a published article.

Last, the Replication Bayes factor uses the effect size estimate from the original study at face value. This is in many cases a questionable assumption, since publication bias and $p$-hacking are known to inflate effect size estimate in the reported literature. For sample size planning it is relevant to account for this, e.g. by planning a study at least about 2.5-times as large as the original study \citep{Simonsohn2015}. For the analysis of a new replication study on an effect which might not yet have been estimated through meta-analyses the comparison with an originally reported effect size is useful and often the first step in an analysis. The Replication Bayes factor can then answer the question: How much more evidence in favor of an effect of this size does the data provide when compared to a null effect? A Bayes factor using a manually (i.e. more subjectively) chosen prior, as mentioned above, would be able to incorporate a corrected effect size. By what factor or procedure one should correct a reported effect size is an open question in meta-analytical research. \cite{GelmanEdlinFactor} has used the term ''Edlin factor'' for this: a measure by which published estimates should be correct by.

\subsection{Conclusion}
The Replication Bayes factor introduced by \cite{Verhagen2014} and extended herein is one index to evaluate the results of a replication attempt. It is, obviously, not able to cover all questions and pitfalls in the analysis of a replication. Instead it is a way to formally and transparently integrate previously available information in the analysis within the Bayesian framework and allows to quantitatively assess the gained evidential value. It is further easy to apply to frequentist results as it uses reported test statistics from the original and replication study only. To cover replications comprehensively, however, researchers have to use different tools depending on the question asked. No single statistical index is sufficient to globally assess the quality of a study or a theory. This is true not only for $p$-values \citep{Wasserstein2016} but also for Bayes factors.

\pagebreak 
\section*{Acknowledgements}
I thank Farid Anvari, André Beauducel, Nicholas Coles, Peder Isager, Anne Scheel, Dani\"el Lakens, Eric-Jan Wagenmakers and one anonymous reviewer for their constructive remarks and comments, that helped to improve the clarity and structure of the manuscript.

\section*{Conflicts of Interest}
None.

\bibliographystyle{apa}
\bibliography{Bibliography-BFrep-ANOVA}

\end{document}